\documentclass[preprint,prd,aps,superscriptaddress]{revtex4}%
\usepackage{psfrag}%12pt
\usepackage{graphicx}%Gráficos

\begin{document}
%\preprint{hep-th/yymmnnn}

\title{Modified Mean Field approximation for the Ising Model}

\author{Cayetano Di Bartolo}
\affiliation { Departamento de F\'{\i}sica, Universidad Sim\'on
Bol\'{\i}var,\\ Aptdo. 89000, Caracas 1080-A, Venezuela.}

\author{Lorenzo Leal}
\affiliation { Departamento de F\'{\i}sica, Universidad Sim\'on
Bol\'{\i}var,\\ Aptdo. 89000, Caracas 1080-A, Venezuela.}
\affiliation {Grupo de Campos y Part\'{\i}culas, Centro de  F\'{\i}sica Te\'orica y Computacional,
Facultad de Ciencias, Universidad Central de Venezuela, AP 47270,
Caracas 1041-A, Venezuela.}

\date{\today}

%\pacs{xx.xx, yy.yy}

\begin{abstract}
We study a modified mean-field approximation for the Ising Model in arbitrary  dimension. Instead of taking a ``central" spin, or a small ``drop" of fluctuating spins coupled to the effective field of their nearest neighbors as in the Mean-Field or the Bethe-Peierls-Weiss methods, we take an infinite chain of fluctuating spins coupled to the mean field of the rest of the lattice. This results in a significative improvement  of the Mean-Field approximation with a small extra effort.
\end{abstract}

\maketitle

The Hamiltonian of the Ising model \cite{Ising} in the presence of an external magnetic field $B$ can be written as

\begin{equation} \label{ising}
 H= - J \sum_{<ij>} s_{i}s_{j} - B\sum_{i}s_{i}.
\end{equation}

\noindent The spins $s_{i}$ can take the values $\pm1$. The first sum is over  pairs of nearest neighboring spins.  The partition function is given by

\begin{equation} \label{particion}
  Z=\sum_{\text{Configurations}} \exp (-\beta H),
\end{equation}

\noindent where $\beta$ is the inverse of the temperature. From now on we absorb the parameter $J$ into $\beta$.

As it is well known, (\ref{particion}) can be exactly summed up only for dimensions $d=1,2$  (the solution of the two dimensional case \cite{Onsager} is far from being trivial). For higher dimensions we must content ourselves with approximate analytical solutions or computer simulations.

The mean-field (MF) approximation \cite{Pierre,Bragg} is a ve\-ry simple and intuitive self-consistent approximation method that predicts some rough features of the Ising model, and of other models that undergo phase transitions. The method consists on ``freezing" all the lattice spins to a fixed value $h$, except one of them (the ``central" spin $s_{0}$). With this simplification, the magnetization per unit of volume $<s_{0}>$ is given by

\begin{equation} \label{magnetizacion}
  <s_{0}>= \sum_{s_{0}=\pm 1} P(s_{0}) \,s_{0} ,
\end{equation}

\noindent where the probability $P(s_{0})$ of finding the  value $s_{0}$ for the central spin is

\begin{equation} \label{pcero}
 P(s_{0}) = \frac{\exp (2d\beta h s_{0})}{\exp (2d\beta h) + \exp (-2d\beta h)
 } ,
\end{equation}

\noindent for the case $B=0$, which is the one we are interested in.

When $<s_{0}>$ is equated to the mean-value $h$, one obtains the transcendental equation

\begin{equation} \label{mf-ecuacion}
 h = \tanh (2dh\beta),
\end{equation}

\noindent that yields  estimates for the magnetization and the critical temperature of the exact model. The MF method predicts a phase transition for $\beta_{MF}=\frac{1}{2d}$, with $h=0$ for $\beta < \beta_{MF}$ and $h\neq0$ for $\beta > \beta_{MF}$. Thus, MF incorrectly yields a phase transition for $d=1$. However, as $d$ increases the results improve: for $d\rightarrow \infty $,\ $\beta_{MF}$ fits the exact result (see table~\ref{Tabla}).

The MF approach can be formulated as a variational approximation as follows. Since the Gibbs probability  distribution minimizes the Helmholtz free energy $\phi= E- \frac{1}{\beta}S$ ($E$ and $S$ are the average energy and the entropy respectively), the probability distribution \textit{ansatz}

\begin{equation}
 P_{\alpha} = \frac{\exp (\alpha \sum_{i} s_{i})}{\sum_{\text{Config.}} \exp (\alpha \sum_{i} s_{i})}
\end{equation}

\noindent will produce a free energy $\phi _{\alpha}$ that overestimates the exact one. Now, varying $\phi _{\alpha}$ with respect to $\alpha$ it is easy to obtain that its minimum value is attained when $\alpha = \tanh (2d\alpha\beta)$, which  just corresponds to the MF \textit{ansatz}. Hence, $\phi_{MF}$ can also be used to estimate the Helmholtz free energy $\phi$. Furthermore, it also makes sense to estimate other thermodynamical observables from this approximate  free energy $\phi_{MF}$.  There also exist MF based approaches, that instead of considering just one fluctuating spin, take a small sublattice (it could be a ``central" spin and its nearest neighbors, for instance) and couple it with the mean field of the background provided by the rest of the lattice \cite{Weiss}, \cite{Bethe}, \cite{Peierls}. These approaches improve the estimations of the critical temperature and the modeling of the physical observables. However, since in essence all of them deal with finite lattices, the estimations of correlation functions are somewhat artificial [by considering an external non-uniform magnetic field and using the linear response theorem  \cite{Binney} one can estimate correlation functions from the MF approximation].

Consider now the following approximation, which we propose as a natural improvement of the MF one that also allows for a simple estimation of correlation functions. Instead of taking a single fluctuating spin, let us consider a straight chain of $N$ ($N\rightarrow\infty$) fluctuating spins in the $d$-dimensional lattice, which are coupled to the constant field $M$ of their remaining ``frozen" nearest neighbors. It is easy to see that this amounts to dealing with the one-dimensional Ising model in an external magnetic field $B=2(d-1)M$.  The probability distribution $P_{l}$ for this line-approximation is given by

\begin{equation} \label{p-mf-modif}
 P_{l} = \frac{\exp\, \left( \beta \sum_{i}^{N} s_{i} s_{i+1} + \beta \sum_{i}^{N}
 2(d-1)M\,s_{i} \right) }{Z_{l}},
\end{equation}

\noindent where the partition function $Z_{l}$ is

\begin{equation} \label{Zl}
 Z_{l} = \sum_{\text{Config.}}\,\exp\, \left( \beta \sum_{i}^{N} s_{i} s_{i+1} + \beta \sum_{i}^{N}
 2(d-1)M\,s_{i} \right) .
\end{equation}

\noindent $Z_{l}$ can be  easily calculated using, for instance, the transfer matrix method \cite{Kramers}. One has

\begin{equation} \label{Transfer}
 Z_{l} = Tr({\hat{T}^{N}}),
\end{equation}

\noindent with

\begin{equation} \label{matriz}
T= \left( \begin{array}{cc}
\exp{(\beta +b)} & \exp{(-\beta)}  \\
\exp{(-\beta)} & \exp{(\beta - b)} \\
\end{array} \right).
\end{equation}

\noindent Here, we have set $b=2(d-1)\beta M$.

The matrix $T$ can be readily diagonalized. Its eigenvalues are

\begin{equation} \label{autovalores}
 \lambda_{\pm} = \exp(\beta) \left( \cosh(b) \pm \sqrt{\sinh^{2}(b) +
 \exp(-4\beta)} \,\right).
\end{equation}

\noindent In the thermodynamical limit ($N \rightarrow \infty$) the partition function only picks the contribution of the greatest eigenvalue. The result is

\begin{equation} \label{Z-explicita}
 Z_{l} = \left[ \exp(\beta) \left( \cosh(b) +\sqrt{\sinh^{2}(b) +
 \exp(-4\beta)}\; \right)\right]^{N}.
\end{equation}

The self-consistency condition $\langle s_{i}\rangle=M$ is now readily seen to be

\begin{equation} \label{consistencia}
M=\frac{1}{2(d-1)\beta N}\, \frac{\partial}{\partial M}\ln Z_{l},
\end{equation}

\noindent which leads to the transcendental equation.

\begin{equation} \label{trans}
M=\frac{\sinh(b)}{\sqrt{\sinh^{2}(b) + \exp(-4\beta)}}.
\end{equation}

\noindent Equation (\ref{trans}) always has $M=0$ as solution. A second solution with a lesser free energy appears for $\beta <\beta_{c}$, where $\beta_{c}$ is given by

\begin{equation} \label{trans1}
2(d-1)\beta_{c} \exp(2\beta_{c})=1.
\end{equation}

\begin{figure}[p]
\psfrag{A}{$d=4$}%
\psfrag{B}{$d=3$}%
\psfrag{C}{$d=2$}%
\psfrag{f}{$M$}%
\psfrag{g}{$\beta$}%
\psfrag{a}{0.2}%
\psfrag{b}{0.4}%
\psfrag{c}{0.6}%
\psfrag{d}{0.8}%
\psfrag{e}{1.0}%
\includegraphics[width=82mm]{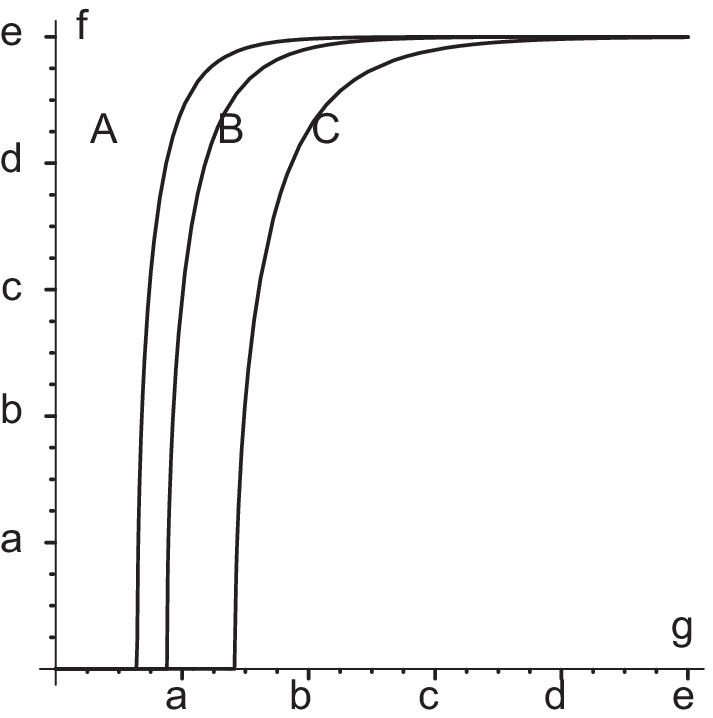}
\caption{Magnetization $M$ versus $\beta$ for several  dimensions.\label{FigMagn}}
\end{figure}

Table~\ref{Tabla} shows the results for several dimensions. Besides the modest improvement  in comparison with ordinary MF theory that this approximation yields for intermediate dimensions, the results for the limiting cases $d=1$ and $d=\infty $ are exact. As in the ordinary MF case, the magnetization $M$ vanishes for $\beta < \beta_{c}$ , while $M\neq0$ for $\beta > \beta_{c}$. In figure~\ref{FigMagn} we show the magnetization vs. $\beta$ for several dimensions. The behavior is qualitatively similar to the ordinary MF case, although the critical temperatures are improved in the present case. From equations (\ref{trans}) and (\ref{trans1}) one can obtain the magnetization near the critical point as

\begin{equation}
M = \sqrt{\frac{6(2\beta_c + 1)}{3\beta_c - \exp(-4\beta_c)}~}\; (\delta \beta)^{1/2} + \text{O}(\delta \beta)^{3/2},
\end{equation}

\noindent with $\delta \beta \equiv \beta -\beta_c \geq 0\,.$ The magnetization critical exponent is $\frac{1}{2}$, just as in the ordinary MF case.

It is worth noticing that the modified MF or line-approximation presented here can also be obtained from a variational principle. To see this we consider the trial probability distribution

\begin{equation} \label{probabilidad-linea}
 \widetilde{P} =\prod_{x-\text{axes}}\exp (\theta \sum_{i=1}^{N} s_{i}s_{i+1} + \alpha \sum_{i}s_{i}),
\end{equation}

\noindent that may be understood as follows. The whole d-dimensional lattice is partitioned into infinite chains parallel to the x-axis. Spins belonging to different chains do not interact. $\theta$ and $\alpha$ are variational parameters, that  must be tuned to produce the best trial probability distribution, i.e., that minimizing the trial Helmholtz free energy $\widetilde{\phi}= \widetilde{E}- \frac{1}{\beta} \widetilde{S}$. Here, $\widetilde{E}$ is the average of the Hamiltonian calculated by using the trial probability distribution, while $\widetilde{S}= - <\ln \widetilde{P}>$ is the entropy associated to that probability distribution. Once again, using the transfer matrix formalism, the corresponding partition function is readily seen to be

\begin{eqnarray} \label{particion-linea}
 \widetilde{Z} &=& \sum_{\text{Config.}} \,\,\prod_{x-\text{axes}}\exp \left( \theta
 \sum_{i=1}^{N} s_{i}s_{i+1} + \alpha \sum_{i}s_{i} \right)\nonumber\\
 &=& \left[ \exp(\theta) \left( \cosh(\alpha) + \sqrt{\sinh^{2}(\alpha) +
 \exp(-2\theta)} \;\right)\right]^{N_{v}},
\end{eqnarray}

\noindent where $N_{v}=N\times$(number of $x$-axes),  is the total number of vertices of the lattice.

The Helmholtz free energy is then given by

\begin{eqnarray} \label{helmholtz-line}
\widetilde{\phi}&=& \widetilde{E}- \frac{1}{\beta}
\widetilde{S}\nonumber\\
  &=& \left( \frac{\theta}{\beta}-1\right) \frac{\partial}{\partial \theta}\ln \widetilde{Z}
- \frac{1}{\beta}\ln \widetilde{Z} \nonumber\\
&&+ \left( \frac{\alpha}{\beta}-
  \frac{(d-1)}{N_{v}}\,\frac{\partial}{\partial \alpha}\ln
  \widetilde{Z}\right) \frac{\partial}{\partial \alpha}\ln
  \widetilde{Z} \,.
\end{eqnarray}

\noindent When the gradient of $\widetilde{\phi}$ with respect to the parameters $(\theta , \alpha)$ is equated to $0$, it is found that there is an absolute minimum for

\begin{eqnarray} \label{parametros}
\theta &=& \beta \\
\alpha &=& \frac{2(d-1)\beta}{N_{v}}\,\,\frac{\partial \ln
\widetilde{Z} }{\partial \alpha},
\end{eqnarray}

\noindent which corresponds precisely to the modified MF solution discussed before. In this case, the Helmholtz free energy per site results to be

\begin{equation}
\widetilde{\phi}/N_v = (d-1) \beta M^2 - \ln(\lambda_+).
\end{equation}

\begin{figure}[p]
\psfrag{A}{A}%
\psfrag{B}{B}%
\psfrag{i}{$\beta$}%
\psfrag{j}{$\beta\widetilde{\phi}/N_v$}%
\psfrag{a}{0.05}%
\psfrag{b}{0.10}%
\psfrag{c}{0.15}%
\psfrag{d}{0.20}%
\psfrag{e}{-0.70}%
\psfrag{f}{-0.71}%
\psfrag{g}{-0.72}%
\psfrag{h}{-0.73}%
\includegraphics[width=82mm]{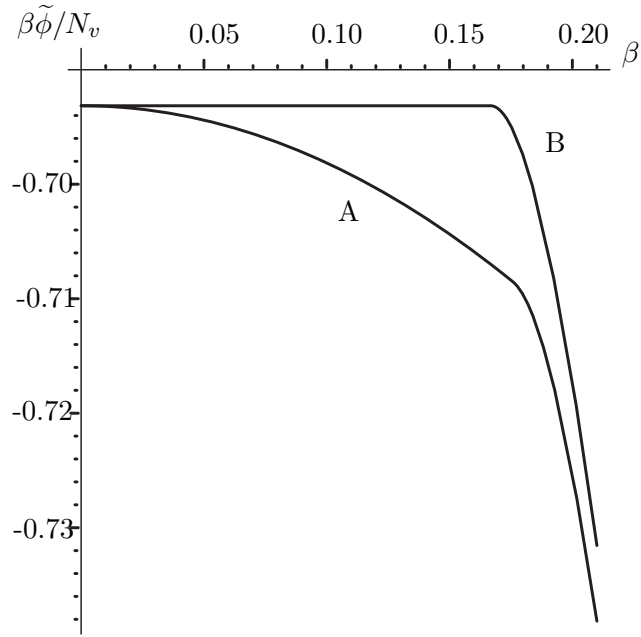}
\caption{Free energy per site (times $\beta$)  versus $\beta$ for $d=3$. Curve A: modified MF approximation, curve B: ordinary MF method. \label{FigFEnergy}}
\end{figure}

In figure~\ref{FigFEnergy} the free energy per site times $\beta$ is depicted for dimension $d=3$ in comparison with the MF one. It is evident that the MF results are improved by using the modified MF approximation that we are presenting here. For other dimensions the results are similar, and the modified MF approximation also improves the ordinary MF ones.

The modified MF method also allows us to estimate  the correlation length $\xi$, employing well known results for the one dimensional Ising model in an external magnetic field. Using the transfer matrix formalism it is easy to see that the connected correlation function $<\sigma_{i} \sigma_{j}>_{c}\equiv <\sigma_{i} \sigma_{j}> - <\sigma_{i}><\sigma_{j}>$ is given exactly by

\begin{equation} \label{función de correlación}
 <\sigma_{i} \sigma_{j}>_{c}= \text{constant} \times \exp \left( - \frac{|j-i|}{\xi} \right) ,
\end{equation}

\noindent where the correlation length $\xi$ is given by

\begin{equation} \label{longitud de correlación}
\xi = \frac{1}{\ln \,\left( \frac{\lambda_{+}}{\displaystyle \lambda_{-}} \right)} \,.
\end{equation}

\begin{figure}[p]
\psfrag{A}{$\xi$}%
\psfrag{B}{$\beta$}%
\psfrag{a}{0.2}%
\psfrag{b}{0.4}%
\psfrag{c}{0.6}%
\psfrag{d}{0.8}%
\psfrag{e}{0.2}%
\psfrag{f}{0.3}%
\psfrag{g}{0.4}%
\psfrag{h}{0.5}%
\includegraphics[width=82mm]{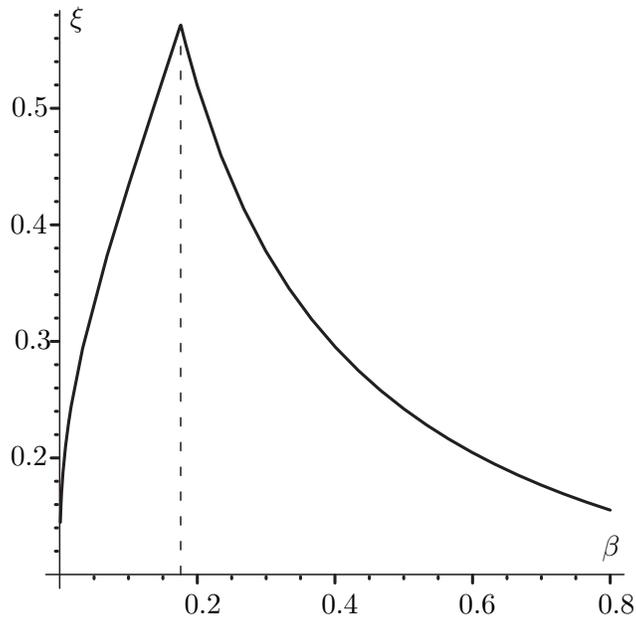}
\caption{Correlation length $\xi$ versus $\beta$ for d=3. The  dashed line corresponds to $\beta_{c}$.\label{FigCorr}}
\end{figure}

\noindent Here, we have taken the spins  $\sigma_{i}$ and $\sigma_{j}$ along the same $x$-axis of ``alive" spins. Otherwise there would not be correlation among them, just as in the ordinary MF approach. In figure~\ref{FigCorr} we show the correlation length as a function of $\beta$ for dimension $d=3$. We observe with interest that $\xi$  exhibits a pronounced peak precisely when $\beta$ reaches the critical value predicted by the model. For other dimensions the graphics  are similar. As it is well known, it is expected that correlation lengths diverge at critical points. So, we find it satisfactory that this simple approximation can mimic that highly non-trivial behavior in a neat way. It should be stressed that this is an analytical result of the approximation we have discussed, i.e. the modified MF approximation, which has no counterpart within the ordinary MF method, since in that case there is no correlation among spins (as we pointed out at the beginning, however, there are ``tricks" to estimate correlation lengths in ordinary MF theory, that rely on putting an external non-uniform magnetic field and employing the linear response theorem).

Summarizing, we have presented a refinement to the MF approximation for the Ising Model that: 1) Improves the estimations of the critical temperature, in comparison with the ordinary MF approximation, 2) fits exactly to the results of the one dimensional case (unlike normal MF) and the infinite dimensional one (as MF also does), 3) improves the estimates for the free energy and, 4) allows to compute correlation functions that exhibit a behavior at the critical temperature that strongly resemble the expected behavior of the underlying physical models. The approximation is entirely analytical, and heavily relies in the exact study of one-dimensional systems.

It could be interesting to apply this approach to other models whose one dimensional cases be resoluble. It also looks reasonable to think in adapting it to the case of lattice gauge theories. In the last case,  the two dimensional models could serve as the starting point to set up the approximation, since gauge-fixing allows to solve exactly the gauge theories in this dimension. These and other interesting issues are under work.

\begin{acknowledgments}
This work was supported by DID-USB grant GID-30 and Fonacit grant G2001000712. We want to thank our colleague Gloria Buend\'{i}a for a useful remark.
\end{acknowledgments}

\begin{table}[p]
\begin{ruledtabular}
\begin{tabular}{|c|c|c|c|}
d & $\beta_{c}$ MF  & $\beta_{c}$ modified MF & $\beta_{c}$ other methods\\
\hline
1 & 0.500 & $\infty$ & ~~$\infty$ \hspace{2.5mm}(exact result)\\[0.5ex]
2 & 0.250 & 0.284 & 0.441 (exact result)\\[0.5ex]
3 & 0.167 & 0.176 & \hspace{1mm}0.222 (Monte Carlo)\\[0.5ex]
4 & 0.125 & 0.129 & \hspace{1mm}0.150 (Monte Carlo)\\[0.5ex]
$\infty$ & $\displaystyle\frac{1}{2d}$ & $\displaystyle\frac{1}{2d}$ & $\displaystyle\frac{1}{2d}$\\[1.5ex]
\end{tabular}
\end{ruledtabular}
\caption{Results for  $\beta_{c}$ compared with ordinary MF and other methods.  \label{Tabla}}
\end{table}


\begin{thebibliography}{99}

\bibitem{Ising}
E.~Ising, {\it Z. Phys.}, 31, 253 (1925).

\bibitem{Onsager}
L.~Onsager, {\it Phys. Rev.}, 65, 117 (1944).

\bibitem{Pierre}
P.~Weiss, {\it J. de Phys.}, 6, 661 (1907).

\bibitem{Bragg}
W.L.~Bragg and E.J.~Williams, {\it Proc. Roy. Soc.}, A, 145 (1934).


\bibitem{Weiss}
P. R.~Weiss, {\it Phys. Rev.}, 74, 1493 (1948).

\bibitem{Bethe}
H.A.~Bethe, {\it  Proc. Roy. Soc.}, 150, 552 (1935).

\bibitem{Peierls}
R.~Peierls, {\it  Proc. Roy. Soc.}, 154, 207 (1936).

\bibitem{Binney}
J.J.~Binney, N.J.~Dowrick, A.J.~Fisher, and M.E.J.~Newman, {\it The Theory of Critical Phe\-no\-me\-na}, Oxford Science Publications (1992).


\bibitem{Kramers}
H.A.~Kramers and G.H.~Wannier, {\it Phys. Rev.}, 60, 252 (1941).


\end{thebibliography}
\end{document}